\documentstyle[12pt]{article}
\def \blankline{\vspace{0.4 cm}}
\def\aprle{\buildrel < \over {_{\sim}}}
\def\aprge{\buildrel > \over {_{\sim}}}

\def \nmed {\langle N_\mu(E_{\rm min},E_0) \rangle}

\begin{document}

\begin{minipage}[t] {4.1 cm}
\begin {flushleft}
{\bf Preprint n.956} \\
July 15, 1993.
\end{flushleft}
\end{minipage}
\begin{minipage}[t] {11 cm}
\begin{flushright}
\begin{minipage}[t] {7 cm}
\begin{center}
Dipartimento di Fisica \\
Universit\'a  di Roma ``la Sapienza'' \\
I.N.F.N. sezione di Roma
\end{center}
\end{minipage}
\end{flushright}
\end{minipage}

\vspace {3.0 cm}

\section*{\centerline {TeV Muons in Hadronic Showers} }

\blankline
\blankline
\centerline {P. Lipari}

\blankline
\centerline {\it I.N.F.N., Sezione di Roma, and}
\centerline {\it Dipartimento di Fisica, Universit\`a di Roma ``la Sapienza",}
\centerline {\it Piazzale A. Moro 2,   I-00185 Roma, Italy}

\vspace {2.0 cm}

\begin {abstract}
{}From the study of   the  multiple muon
events in deep underground detectors, it is possible
to extract information
about the  spectrum and composition of the primary cosmic rays.
In this work  the   number of TeV  muons
produced by a primary cosmic ray  of a given energy and
zenith angle is computed using analytic  and montecarlo methods,  for
a family of simplified models  as description of the properties
of hadronic interactions.
The  effects of the  uncertainties
in our knowledge of the hadronic cross sections
in the calculation of TeV muons are discussed.
\end{abstract}

\newpage

\section {Introduction}
Deep underground detectors measure  events with several
nearly parallel muons \cite{Frejus,NUSEX,MACRO-composition},
separated by few meters
\cite{MACRO-decoherence}. The
muons  are produced  by  the decay of charged pions and kaons
in the hadronic  showers induced by primary cosmic rays.
{}From the study of these events
it is possible to obtain information
about the spectrum and composition of cosmic rays
in the energy region ($10^{14} \aprle E_0 \aprle  10^{17}$~eV).

The shower produced by a nucleus of
total energy $E_0$ and mass number $A$ can be described
with  reasonable accuracy as
the superposition of $A$ proton showers each of energy
$E_0/A$.
Therefore
if a proton of energy $E_0$  produces
(above a threshold energy  $E_{\rm min}$, and
with a zenith angle dependence $\propto (\cos \theta)^{-1}$)
an average number of
muons   $\langle N_\mu (E_0) \rangle_p$,
a nucleus
of mass number $A$  and the same total energy
will produce with good approximation \cite{Gaisser,semi-superposition}
an average  number of muons:
$\langle N_\mu (E_0) \rangle_A \simeq A \langle N_\mu(E_0/A) \rangle_p$.
The average muon multiplicity of a proton shower grows
more slowly that the  primary energy, approximately as
$E^{0.75}$;  this  can be understood  observing that  in the approximation
of Feynamn scaling the number of  mesons above energy $E_{\rm min}$
in a shower, neglecting threshold effects, grows linearly with  energy,
but with increasing energy the mesons are produced deeper
in the atmosphere, where their decay is more  rare.
Therefore the average number of  high energy muons
in a shower of  energy $E_0$
depends on the mass number of the primary nucleus approximately as
$\propto A^{0.25}$, and (for the same energy spectrum)
a primary comic ray flux  rich in heavy nuclei
will produce more high multiplicity
events in deep undergroud muon detectors
than a flux with a `light' composition.
This allows us in principle to
obtain information  on the primary cosmic ray spectrum and composition
from underground measurements of multiple muon events.
The uncertainty on the properties of the hadronic interactions
is  the main   source of systematic errors in the development of
this program.

In this work we will discuss some methods to estimate the size
of this uncertainty.
Elbert \cite{Elbert}  has suggested
that the average number  of muons above a minimum energy
$E_{\rm min}$ produced in the shower of a primary
proton of energy $E_0$
has the scaling form:
\begin{equation}
\nmed = {1 \over E_{\rm min} ~\cos \theta} ~
 G \left ( {E_{\rm min} \over E_0 } \right ) ,
\label {G-scal}
  \end{equation}
Gaisser and Stanev \cite{NIM85}
and  Forti {\it et al.} \cite{Forti} have also fitted their montecarlo
results with the form (\ref{G-scal}).
In section 2 we will derive  formally
this result, discussing under which
conditions it is valid, we will also show
how it is  possible to compute  with analytic methods the  function $G(x)$
from a knowledge  of the inclusive single--particle
differential cross sections.

In section 3 we will  discuss a family of  `toy models' for the hadronic cross
sections that are a generalization of the algorithms
introduced by Hillas \cite{Hillas}.
These models are fully described by a small
set of simple montecarlo algorithms, and are therefore  constructed
as montecarlo  instruments, at the same time
the  inclusive differential cross sections  generated by the
algorithms   have simple
closed  form analytic representations, and are also suitable for
analytic studies.

In section 4  we will discuss some explicit calculations
of  high energy muon production in proton showers,
comparing    analytic and montecarlo  methods.
The montecarlo technique is necessary for a correct treatment of
fluctuations in the shower development.

In section 5 we will  compare two models of hadronic interactions
that are constructed to  produce the same inclusive muon flux,
with one model having a larger and softer $\pi$ multiplicity, and
discuss how the multiplicity distribution of TeV muons   changes.

\section {Analytic Method}
In this section we will discuss  how under two approximations:
(i) energy independent interaction lengths,
(ii) validity
of Feynman scaling  in the fragmentation region of the hadronic cross sections,
it is possible to solve analytically  the  shower development equations
and calculate  the
inclusive high energy
muon spectrum  produced by  a primary  proton of energy $E_0$.
The solution  has the scaling form (\ref{G-scal})
suggested by
Elbert \cite{Elbert} and used in \cite{NIM85,Forti}, (there is also
 a differential form of this scaling law given in equation (\ref{Fdef})).
The function $G(x)$  depends on the
interactions lengths of  nucleons, and mesons,
and on the inclusive differential cross sections
$d\sigma_{a \to b}/dx$  for production of particle $a$ in the interaction
of particle $b$ with $x = E_b/E_a$.
There is also a dependence on the density distribution
of the medium where the showers develop. For an exponential
distribution: $\rho (h) \propto e^{-h/h_0}$,
as is the case  for  cosmic rays reaching the  earth atmosphere
and  zenith angles not too close to the horizontal direction,
this results in $G$ being proportional  to the scale height
$h_0$.

The  analytic calculation of  the functions $G$
can be performed with a straightforward generalization of the methods
used  for the   calculation of  electromagnetic showers \cite{Rossi}.
We will begin to discuss the inclusive muon flux
produced  by a power law flux of primary particles, and then discuss
the case of a monochromatic  beam of particles.

\subsection {Power law initial proton spectrum}
The calculation of the inclusive  muon flux
produced by a power law primary flux,   is
a well known problem (see for example the textbook \cite{Gaisser}),
and is directly applicable to the real flux  of primary particle
reaching the earth.
The all nucleon cosmic rays spectrum
is in fact  well  represented by  a power law:
$\phi_0 (E_0) \simeq K ~E_0^{-\alpha}$ with index $\alpha = 2.7$, and
a constant $K \simeq  1.85$
(with $E_0$ in GeV and $\phi$ in (cm$^2$~s~sr~GeV)$^{-1}$).
Using the approximations  of constant interactions lengths and
Feynman scaling,  the
muon spectrum  above an energy of $\sim 20$ GeV
has the approximate form \cite{Gaisser}:
\begin{equation}
\phi_\mu (E, \theta)  = \left \{
L_\pi(\alpha) ~ \left [ 1 + {L_\pi(\alpha) \over H_\pi (\alpha)}
 {E \cos \theta \over \varepsilon_\pi} \right]^{-1}
 +
L_K (\alpha) ~ \left [ 1 +
 {L_K (\alpha) \over H_K (\alpha)}
{E \cos \theta \over \varepsilon_K} \right]^{-1}
\right \} ~K ~E^{-\alpha}
\label {fluxmu}
\end{equation}
We will  be mostly interested in  the   high energy limit
($E \gg \varepsilon_\pi, \varepsilon_K$):
\begin{equation}
\phi_\mu(E, \theta) = {\varepsilon_\pi~H_\pi(\alpha)
 + \varepsilon_K~ H_K(\alpha) \over
E ~\cos\theta} ~ K~E^{-\alpha} =
{\epsilon_\mu (\alpha)
\over E ~\cos\theta} ~ K~E^{-\alpha}
\label{fluxh}
\end{equation}
The critical energy $\varepsilon_\pi = m_\pi h_0/ c \tau_\pi \simeq 115$ GeV
can be interpreted approximately as the energy at wich  the
average charged pion
produced in a shower has a probability of 1/2 to decay,
$\varepsilon_K \simeq 850$ GeV has the same meaning for charged kaons,
$m_\pi$ ($m_K$)  and $\tau_\pi$  ($\tau_K$)
are the mass  and lifetime of a charged pion (kaon),
$h_0 = 6.34$ km is the scale--height of the stratosphere.
The  `constants' $L_\pi$, $L_K$, $H_\pi$ and $H_K$, depend
on the esponent of the primary flux $\alpha$, and on the properties
of hadronic interactions.
Explicit  solutions  for
$L_\pi(\alpha)$ and $H_\pi(\alpha)$  are
\begin{equation}
L_\pi (\alpha) =
Z_{N\pi}
{}~\left [ {1 - r_\pi^{\alpha} \over \alpha (1- r_\pi)} \right ]
\label {Lpi}
\end{equation}
\begin{equation}
H_\pi (\alpha) =
{Z_{N\pi} \over 1 - Z_{NN}}
{}~{\Lambda_\pi \over \Lambda_\pi - \Lambda_N}
\ln \left ( {\Lambda_\pi \over \Lambda_N} \right )
{}~\left [ {1 - r_\pi^{\alpha+1} \over (\alpha+1)(1- r_\pi)} \right ]
\label {Hpi}
\end{equation}
In these equations: $r_\pi = (m_\mu/m_\pi)^2$;
the  quantities $Z_{ab} (\alpha)$ are defined as:
\begin{equation}
Z_{ab}(\alpha) = \int_0^1 dx ~x^{\alpha-1}
{}~\left ( {dn(x) \over dx} \right )_{a \to b},
\end{equation}
where $dn_{a \to b}/dx$ is the   spectrum of particle $b$ in the
interaction of particle $a$
with a target   air nucleus and  $x= E_b/E_a$;
$\Lambda_N = \lambda_N/(1 - Z_{NN})$,
$\Lambda_\pi = \lambda_\pi /(1 - Z_{\pi\pi})$
where $\lambda_N$ and $\lambda_\pi$ are the interactions lengths in
air for nucleons and pions.
The constants $L_K(\alpha)$ and $H_K(\alpha)$ have
similar expressions, the inclusion of
the processes ($\pi^\pm \to K^\pm$) and
($K_L \to K^\pm$)  introduces some small complications \cite{lip-lep}.
The flux of only positive or negative muons  can also be calculated
with expressions similar to (\ref{fluxmu}).

\subsection {Monochromatic initial proton spectrum}
The flux of muon produced  by a
monochromatic beam of protons takes a more complicated  form.
We can define the function $f(E;E_0)$, the  inclusive
differential spectrum of  $\mu$'s
produced by a primary proton  of energy $E_0$,
and $g(E_{\rm min}; E_0)$ the integral
spectrum
(or average number of muons) above energy $E_{\rm min}$.
The functions $f$ and $g$ are related by:
\begin{equation}
g(E_{\rm min}; E_0) \equiv \nmed = \int_{E_{\rm min}}^{E_0} dE ~f(E; E_0)
\label {fgrel}
\end {equation}
If $E$  ($E_{\rm min}) \gg \varepsilon_\pi, \varepsilon_K$,
equation (\ref{fluxh}) implies:
\begin{equation}
\left [ \int_E^\infty dE_0~E_0^{-\alpha} ~f(E, E_0) \right ]~E^{\alpha+1} =
\epsilon_\mu  (\alpha) \equiv
\varepsilon_\pi ~H_\pi(\alpha) +
\varepsilon_K ~H_K(\alpha)
\label{step-a}
\end{equation}
This condition can be satisfied only if the
functions $f$ and $g$ have the
scaling form :
\begin{equation}
f(E;  E_0)= {1 \over \cos \theta}
{1 \over E^2 } F \left ({E \over E_0 } \right )
\label {Fdef}
\end {equation}
\begin{equation}
g(E_{\rm min}; E_0)=
{1 \over \cos \theta}
{1 \over E_{\rm min}} G \left ({E_{\rm min} \over E_0 } \right )
\label {Gdef}
\end {equation}
Equation (\ref{step-a})  defines also
the Mellin transforms $M_F$ and $M_G$ of the function $F$ and $G$:
\begin{equation}
M_F(s) = \epsilon_\mu(s+2)
\label{F-tra}
\end{equation}
\begin{equation}
M_G(s) = { \epsilon_\mu (s+2) \over s+4}
\label{G-tra}
\end{equation}
The Mellin transform of a functions $A(x)$
defined in the interval $0 \le x \le 1$ is:
\begin{equation}
M_A (s) = \int_0^1 dx~ x^s A(x)
\label {Mellin-def}
\end{equation}
{}From
equation (\ref{fgrel})  it is possible  to deduce the
following relation between the functions $F$ and $G$ :
\begin{equation}
F(x) = -x^2 {d \over dx} \left [ {G(x) \over x} \right ] .
\label{FG-relation}
\end{equation}
that implies in general $M_G(s) = M_F(s)/(s+4)$.

For continuous $A(x)$
the  transform $M_A(s)$ is well defined  in the complex field
for all  values of the imaginary part of its argument
 $\Im[s]$,  and for the  real part
in the interval
$s_1 \le \Re[s] \le s_2$.
The relation (\ref{Mellin-def}) can be inverted \cite{mellin,Rossi} with:
\begin{equation}
A(x) = {1 \over 2 \pi i} \int_C ds~ M_A(s) x^{-i(s+1)}
\label{mell-inv}
\end{equation}
where the integral  is taken   in the complex field along an
arbitrary  path $C$  that starts from
$\Im(s) = -\infty$  and ends at $\Im(s)=+\infty$,
and is entirely contained in
the domain  of definition of $M_A(s)$.

To summarize:  the function $\epsilon_\mu(\alpha)$,
introduced in (\ref{fluxh}), for $\alpha$ real  and   positive
determines  the high energy inclusive muon flux
produced by a power law primary  flux of index $\alpha$:
 $\phi_\mu(E)/\phi_0(E) = \epsilon_\mu(\alpha)/(E~ \cos \theta)$.
The function $\epsilon_\mu(\alpha)$   can be
calculated from the inclusive hadronic cross section
according to equations (\ref{fluxh}) and (\ref{Hpi}),
and the definition  is  valid also
for $\alpha$ in an infinite region of the complex field.
According to equations (\ref{F-tra}) and  (\ref{G-tra}),
$\epsilon_\mu(\alpha)$
gives also the Mellin transform of the
functions $F(x)$ and $G(x)$ that
determine the  differential and integral inclusive muon
spectrum for a monochromatic  beam of protons.
We can therefore compute the functions $G(x)$ and $F(x)$ in two steps:
(i) compute the function $\epsilon_\mu(\alpha)$, (ii) use the inversion formula
(\ref{mell-inv}).

\section {Models for the  differential cross sections}
In this section we will discuss an explicit family of models
for the  hadronic cross sections, that  is a
simple generalization of the
montecarlo algorithms
proposed by Hillas \cite {Hillas}, to model multiple particle
production in high energy hadronic interactions.
These algorithms, in spite of their simplicity,
give results in good agreement  with  data.
The inclusive differential cross sections  generated
by the   montecarlo algorithms
have also explicit and exact analytic representations,
this allows to cross check results obtained with analytic  and
montecarlo methods.

To explore the sensitivity of  muon production
to the  characteristics of the cross sections, we
have   generalised the set of algorithms originally proposed
by Hillas, introducing   some parameters that can be changed  continuously.
In chapter 17 of reference \cite {Gaisser}, Gaisser describes in some
detail  the Hillas algorithms, and discuss also possible generalizations.
In \cite {Hillas} Hillas
considered only nucleons
and pions, neglecting kaons
and other particles. It is rather straightforward
in the framework of the model
to   include kaons or  treat
separately protons and neutrons,
but for the sake of simplicity
we will  also consider only   nucleons and pions.
The separate treatment of protons and neutrons is  of little importance
if we  sum over the charge of the muons;
only $\sim 25$\% of TeV  muons are produced by kaons, and therefore,
for our illustrative discussion, kaons are not of
crucial importance.
The main focus of this work is not to obtain an absolute calculation
but to discuss the sensitivity of the results to variations
of the `input'.
The algorithms that we are considering are exactly scaling
in the variable $x = E/E_0$ where  $E_0$ is the incident particle
energy. Scaling violations can be be introduced allowing the
parameters of the model to vary with energy, we will however
discuss here only scaling models,
and concentrate  on the study of the effects of  distortions of the spectra
of  nucleons and pions.
We will not  discuss the transverse momentum   of the particles produced
in the shower, and we will not comment on the separation between muons.

\subsection {Proton interactions}
The set of algorithms to generate   the interaction of a proton
of initial energyy $E_0$ are:

\begin {enumerate}
\item  Generate $x$  ($0 \le x \le 1$)  with linear distribution
of average $\langle x \rangle = 1-K_p$. Construct
a leading nucleon of energy $E_0 x$.

\item The remaining energy  $E_0 (1-x)$ is divided into two parts  $A$ and $B$
with a flat distribution.
Then each piece of energy is split  again into two parts
with a  flat distribution ($A \to A_{1} + A_{2}$ and
$B \to B_{1} + B_{2}$).
\item Each of the four pieces
 of energy has now a probability $P^*$ of splitting
once again into  two parts, always  with flat distribution.
\item At this point we have   $N$  energy pieces
($4 \le N \le 8$), with  a binomial probability
distribution $p(N) = P_{\rm binomial}(N-4;~4,P^*)$.
Each  piece of energy now enters
a  recursive process.
\item The energy of the piece is splitted  with flat distribution
into two parts. A part is assigned as the energy of a pion
(with equal probability for the three charges),
the  remaining energy is again  divided into two parts: one is assigned
to a  new  pion,   for the  other the process  is iterated.
The  splitting is stopped when the remaining  energy is  less
than a pion mass or of some preassigned threshold value.
\end{enumerate}

In the algorithms  we have introduced two parameters:
the inelasticity $K_p$ ($1/2 \le K_p \le 2/3$)
that gives the fraction of the projectile particle
energy that  is not transfered to the
leading nucleon,
 and $P^*$
($0 \le P^* \le 1$)
that determines the shape of the  pion spectrum,
when ($x \to 1$) the pion spectrum has the behaviour
$\propto (1-x)^3$
if $P^*=0$ and
$\propto (1-x)^4$
if $P^*=1$.
A `model'  in this framework is  then represented by the pair
of numbers $\{K_p,P^*\}$.
The original algorithms of Hillas
corresponds to the model $\{ {1\over 2}, 0 \}$.

The  montecarlo algorithms described above
give inclusive differential cross sections that have
explicit analytic expressions:
\begin{equation}
{dn_{pp} \over dx} = 1 + 6 \left ( K_p - {1 \over 2} \right )
- 12 \left ( K_p - {1 \over 2} \right ) ~x,
\label{proton-proton}
\end{equation}
\begin{eqnarray}
{dn_{p\pi^\pm} \over dx} & = & {8\over 3}
 \left [ 1- 6 \left ( K_p - {1 \over 2} \right ) \right ]
\left \{ \left [ {1 \over x} -1 + \ln x - { (\ln x)^2 \over 2} \right ]
 ~(1 + P^*) + 2 P^* {(\log x)^3 \over 6} \right \} + \nonumber \\
& + & {96 \over 3} \left ( K_p - {1 \over 2} \right )
\left [ \left ( {1 \over 2x} -{x \over 2}  + \ln x \right) (1 - P^*)
+ 2P^* \left ({1 \over 2x}- 1 + {x \over 2}
 - { (\ln x)^2 \over 2} \right ) \right ]~~~~
\label{proton-pion}
\end{eqnarray}
The differential cross section for $\pi^\circ$ production
is simply 1/2 of  (\ref{proton-pion}).
The   plateau of the rapidity  distribution for the
charged pions has a height  that increases linearly with $(1+P^*)$:
\begin{equation}
\rho_{p \to \pi^\pm} \equiv \left ( {dn_{p \pi^\pm} \over dy}\right )_{y^* = 0}
= {8\over 3}  [1 + P^*]
\label {rho-proton}
\end{equation}
The momenta of the inclusive spectra $Z_{pX}(\alpha)
= \langle x^{\alpha-1} \rangle_{pX}$   can be calculated explicitely:
\begin{equation}
Z_{pp} (\alpha) =
\left [1 + 6 \left ( K_p - {1 \over 2} \right ) \right ] {1 \over \alpha}
- 12 \left ( K_p - {1 \over 2} \right ) {1 \over \alpha +1} ,
\label{z-pp}
\end{equation}
\begin{equation}
Z_{p\pi^\pm} (\alpha) = {8\over 3}
\left [ 1 - P^* \left ( {\alpha -2 \over \alpha} \right ) \right ]
{}~\left [ {1 \over \alpha^3 (\alpha -1) } + \left (K_p - {1 \over 2} \right )
{1 \over \alpha^2} \left ( {12 \over \alpha^2 -1} - {6 \over \alpha }
\right ) \right ]
\label{z-ppi}
\end{equation}
For $\alpha=2$,
$Z_{pp} = 1-K_p$ and $Z_{p\pi^\pm} = {2\over 3} K_p$
independently from  $P^*$ and the shape of the spectra.

\subsection {Charged pion interactions}
To generate a $\pi^\pm$ interaction the algorithms are the following
\begin{enumerate}
\item The total energy  is divided into two parts $A$ and $B$
with a flat distribution. The first part $A$  is assigned to
a pion with  probability $P_D$. This `diffractive' pion is charged with a
 probability of 0.87,  the probability  of being of
same (opposite) sign with respect to the    projectile
is 0.80 (0.07).
\item The energy of the piece $B$
 is divided into two parts $B_1$ and $B_2$, then
each  piece is again divided into two. We now have  4 energy pieces.
$B_{11}$, $B_{12}$, $B_{21}$ and $B_{22}$.
\item Piece $A$
if it was not already assigned to the
diffractive   pion, is treated in the same way as piece $B$ in the previous
and following steps.

\item  The   pieces $B_{11}$ and $B_{21}$ have now  each a probability $P_A$ of
being assigned to a pion (with equal probability for the three
charges).
\item     Each of the remaining energy pieces   ($B_{12}$, $B_{22}$ and
$B_{11}$, $B_{21}$  if not already assigned to  pions) are
splitted into two parts with flat distribution with a probability
$P_B$
\item Part $B$ is now divided into $N$  ($2 \le N \le 8$) energy pieces.
Each of these  pieces is fragmented into particles, with a recursive  process
(as in proton interactions).
The energy of one piece is splitted  with flat distribution
into two parts,  a part is assigned as the energy of a pion,
for the remaining  energy
the process is
iterated  until we are left with  a remaining energy less
than a pion mass or a preassigned threshold value.
\end{enumerate}
The pion interaction model  depends on three parameters
$\{P_D,P_A,P_B\}$.
The original algorithm proposed by  Hillas \cite{Hillas} corresponds to the
choice $\{ {1\over 2} ,{1 \over 2} ,0 \}$.
The physical meaning of the parameters is  easy to  understand intuitively.
Increasing $P_D$ and $P_A$  the pion spectrum  is hardened and the
multiplicity decreased. Increasing  $P_B$  has the opposite effect:
the  spectrum  is
softened, and the multiplicity increases.

The inclusive pion spectrum implied by this set of algorithms
can be  calculated:
\begin{eqnarray}
{dn_{\pi^\pm\pi^\pm} \over dx} & = & 0.87~P_D
 +  {4\over 3} (2-P_D)  P_A { (\ln x)^2 \over 2} +
\label{pion-pion}
\\
& + & {4\over 3} (2-P_A)(2-P_D)
{}~\left [ \left ({1 \over x} -1 + \ln x - { (\ln x)^2 \over 2} \right)
{}~(1 + P_B) +
2 P_B { (\ln x)^3 \over 6}  \right ] \nonumber
\end{eqnarray}
The spectrum of $\pi^\circ$ is obtained changing the numerical coefficients
of the three  terms in (\ref{pion-pion}):  $0.87 \to 0.13$, $4/3 \to 2/3$, and
$4/3 \to 2/3$.
The momenta of the  energy distribution
are:
\begin{equation}
Z_{\pi\pi^\pm} =
{0.87~P_D  \over \alpha} +
{4\over 3} (2-P_D)  P_A { 1\over \alpha^3}
+ {4\over 3} (2-P_A)(2-P_D) { 1 \over \alpha^3(\alpha-1)}
{}~\left [ 1 + {2 P_B \over \alpha} \right ]
\label{z-pipi}
\end{equation}
The rapidity density is:
\begin{equation}
\rho_{\pi\pi} = {4 \over 3} (2 -P_A) (2-P_D)(1+P_B)
\end{equation}

\section {Explicit Calculations}
In this section we will discuss some  calculations of  TeV muons,
using   the   algorithms discussed in the previous section to
describe hadronic interactions.
We will always  assume  that the interactions length are constant
and that Feynman scaling is valid.  In  our framework therefore
an `hadronic interaction model'  is fully described as  a
`vector' of eight numbers:
$M = \{ \lambda_p, \lambda_\pi; ~K_p, P^*; ~P_D,P_A,P_B\}$,
the first  two quantities
are  the  interactions lengths  of nucleons and pions,
the next two numbers  describe
protons  interactions,
the last three numbers  describe
pion interactions.

The first  quantity that is interesting to calculate is the
inclusive high energy muon flux.
This flux is determined by the quantity $\epsilon_\mu(\alpha)$ for
$\alpha=2.7$.
In table 1 we show the  value of $\epsilon_\mu(2.7)$
calculated  for different `hadronic interaction models'.

\vspace {0.5 cm}
\noindent {\bf Table 1.} $\epsilon_\mu(\alpha=2.7)$.
The primary flux has form: $\phi_0(E) = K~E^{-2.7}$;
the muon flux  ($E \gg 100 $ GeV) has form:
$\phi_\mu(E,\theta) = \epsilon_\mu/(E~\cos \theta)
\times \phi_0 (E)$.

\vspace{0.4 cm}
\begin {tabular} {| c | c || c | c || c | c | c || c |}
\hline
$\lambda_p$ (g cm$^{-2}$) &
$\lambda_\pi$ (g cm$^{-2}$) &
$K_p$ & $P^*$ & $P_D$ & $P_A$ & $P_B$ & $\epsilon_\mu$ (GeV) \\
\hline
\hline
 86.0  &   111.8 &    1/2 &    0 &   1/2 &   1/2 &    0 &     8.57 \\
\hline
 77.4  &   111.8 &    1/2 &    0 &   1/2 &   1/2 &    0 &     9.02 \\
 86.0  &   100.6 &    1/2 &    0 &   1/2 &   1/2 &    0 &     8.14 \\
 77.4  &   100.6 &    1/2 &    0 &   1/2 &   1/2 &    0 &     8.57 \\
\hline
 86.0  &   111.8 &     1/3 &    0 &   1/2 &   1/2 &    0 &       5.42 \\
 86.0  &   111.8 &     2/3 &    0 &   1/2 &   1/2 &    0 &      11.03 \\
 86.0  &   111.8 &     1/2 &    1 &   1/2 &   1/2 &    0 &       6.35 \\
\hline
 86.0  &   111.8 &     1/2 &    0 &    0 &   1/2 &    0 &       7.96 \\
 86.0  &   111.8 &     1/2 &    0 &    1 &   1/2 &    0 &       9.34 \\
 86.0  &   111.8 &     1/2 &    0 &   1/2 &    0 &    0 &       8.45 \\
 86.0  &   111.8 &     1/2 &    0 &   1/2 &    1 &    0 &       8.70 \\
 86.0  &   111.8 &     1/2 &    0 &   1/2 &   1/2 &    1 &      8.44 \\
\hline
\end{tabular}

\vspace {0.5 cm}
The first line  of table 1 corresponds to
the  differential cross sections proposed  originally by Hillas
\cite{Hillas}, and  is also our choice for a `reference model',
the other lines show the sensitivity to 10\% changements in the interactions
length, and to   the maximum allowed variations of each of the
parameters that we have constructed to describe the shape
of the differential cross section.
We can observe that the  inclusive muon production is
especially sensitive to the properties of the
proton interactions.
Decreasing the inelasticity $K_p$ or softening the pion spectrum (increasing
$P^*$) depresses the muon  flux.
Modifications of the pion differential cross section
have a  smaller effect on the inclusive muon flux.
In fact, because of the steepness
of the primary flux, most of the  muons are produced
in the decay of   mesons produced in the first interaction
of a  primary cosmic ray.

The function $G(x)$ that  gives the average
number  of high energy muons produced by a primary  particle
is easily calculated from (\ref{G-tra})
using  the inversion formula
(\ref{mell-inv}).
In  the  models we are considering
$\epsilon_\mu(\alpha)$ can be obtained from
(\ref{fluxh}), (\ref{Hpi}), (\ref{z-pp}), (\ref{z-ppi}), (\ref{z-pipi}),
as a simple combination of elementary functions,  and
the Mellin transform  can be inverted with an easy
numerical integration.

The function $G(x)$ calculated as discussed above for the `reference model':
[ $\lambda_p$ = 86
and $\lambda_\pi=111.8$ g~cm$^{-2}$,
$\{K_p, P^*;~P_D,P_A, P_B\} =
\{ {1\over 2}, 0; ~{1\over 2}, {1 \over 2}, 0\}$],
is shown in figure 1.
In the same figure we also show for comparison the
curves that Gaisser and Stanev \cite{NIM85} and Forti {\it et al.}
\cite {Forti} have   used as fit to their montecarlo calculations.
The agreement is surprisingly good considering the
extreme simplicity of the  model we are discussing.
It should also be noted that
the results of \cite {NIM85, Forti} refer to muons
not  above a fixed threshold energy, but at a fixed depth
$h$.  For the comparison
we have used the approximation $E_{\rm min}(h) \simeq 0.53 ~(e^{0.4h}-1)$
($h$ in km.w.e., $E$ in TeV), fluctuations
in the muon energy loss   \cite {LS91} should however be taken into account
for a more detailed comparison.

In figures 2,3,4,5  we  show
how the function $G(x)$ is modified because of
changements in the shape and normalization
of the  inclusive differential cross sections.
We have recalculated  $G(x)$ using different values of the parameters.
We will consider the `reference model',
and change one parameter  at the time.

In figure 2a we show the
($p \to \pi^\pm$) spectrum
obtained with different values
of the inelasticity $K_p = {1 \over 2}$, ${1 \over 3}$,  ${2 \over 3}$
and $P^* = 0$.
In figure 2b, we show the function   $G(x)$ as a function of
 $1/x = E_0/E_{\rm min}$
for these three values of $K_p$. As expected the number of muons
increases for  larger $K_p$, because more energy is    transfered to pions.
With increasing $E_0/E_{\rm min}$ however the difference becomes smaller,
the curves join, and in fact on close inspection cross each other.
This can be explained  observing that   for large $E_0/E_{\rm min}$
it is possible to obtain more pions  above the threshold energy $E_{\rm min}$
giving more energy  to the leading nucleon in the interaction.
The number of pions in the first interaction of the shower decreases, but
additional pions are created in the second interaction,
our calculation takes into account the fact that
these additional pions are produced deeper in the atmosphere
and have a smaller probability to decay.

In figure 3a we show the ($p  \to \pi^\pm)$ spectrum
for $P^*=1, 0$  ($K_p = {1 \over 2}$).
As discussed before   $P^* = 1$ corresponds to a softer spectrum, with a higher
multiplicity.
In figure 3b we show  the function $G(x)$  as a function of
$1/x = E_0/E_{\rm min}$  calculated for the
two pion spectra.
For small $E_0/E_{\rm min}$  the curve
corresponding to $P^*=0$ (harder $\pi$ spectrum)     is higher;
the   two curves
cross each other  at $E_0/E_{\rm min} \simeq 30$.
This is qualitatively easy to  understand:
for small $E_0/E_{\rm min}$
the harder spectrum produces more pions above threshold,
when $E_0/E_{\rm min}$ grows, the effect of the larger pion multiplicity
becomes dominant.

In figure 4a  and 5a we show the  $(\pi^\pm \to \pi^\pm)$ spectrum
for  different  values of the parameters :
$P_D=0, {1\over2}, 1$, and $P_B=0,1$.
In figure 4b and 5b we show the  curve $G(x)$
 calculated with the different spectra.
Some remarks are :
(i)  very large deformation of the $(\pi^\pm \to \pi^\pm)$  spectrum
result in small  variations of
$\nmed$; (ii) the effect  is  very small
for $E_0/E_{\rm min} \aprle 10$   when most muons come from the decay of
first generation mesons; (iii) the  largest effects comes
from different treatments of  $\pi$ diffraction  (fig 4b);
 (iv) a softening  of the  $\pi \to \pi$ spectrum from
$dn/dx \propto (1-x)^3$ ($P_B=0$) to $\propto (1-x)^4$
($P_B=1$)
depresses  muon production
for $E_0/E_{\rm min} \aprle 10^{3}$, then enhances it.
The effect (see fig 5b) is however small.

\vspace {0.5 cm}
To summarize the  information of the effect of $G(x)$ of the
changement of  the parameters of the model,
it can be useful to discuss the quantity $\xi_P (E_{\rm min}/E_0)$,
the logarithm derivative  of the average muon multiplicity
as a function of the  parameter $P$
taken  from the `starting model' that we take as the original Hillas model:
$\{K_p,P^*;P_D,P_A, P_B\}= \{{1 \over 2},0,{1\over 2}, {1\over2}, 0\}$
with $\lambda_p = 86$ and $\lambda_\pi = 111.8$ g~cm$^{-2}$.
\begin{equation}
\xi_P (E_{\rm min}/E_0) =
  {\partial \log \langle N_\mu(E_{\rm min}, E_0) \rangle \over \partial \log P
}
\simeq
  {\Delta \langle N_\mu \rangle \over  \langle N_\mu \rangle }
{}~\left (   {\Delta P  \over  P }\right )^{-1}
\end{equation}
The meaning of
$\xi_P$ is that if the parameter $P$  is changed by (for example)
10\%, the resulting percentual effect on
$\nmed$  is $(\xi_P \times 10)\%$.
A positive (negative)
$\xi_P$ indicates that an increase of parameter
$P$  will produce a depression (enhancement) of
$\langle N_\mu\rangle$.
The logarithimic derivatives $\xi_P(x)$ are shown in table 2.

\vspace {0.5 cm}
\noindent {\bf Table 2.} Logarithmic derivatives $\xi_P$(x).
($x = E_{\rm min}/E_0$).

\vspace{0.4 cm}
\begin {tabular} {| c || r | r | r | r | r |}
\hline
Parameter / $x$ & $10^{-1}$ & $10^{-2}$ &  $10^{-3}$ &  $10^{-4}$ &  $10^{-5}$
\\
\hline
$\lambda_p$   & $-$0.530  &  $-$0.497    & $-$0.411  &  $-$0.333  &  $-$0.276\\
$\lambda_\pi$ &  0.512  &   0.483    &  0.405  &   0.333  &   0.280\\
$K_p$         &  0.985  &   0.230    &  0.082  &   0.033  &  $-$0.002\\
$(1-P^*)$     & $-$0.282  &   0.133    &  0.192  &   0.173  &   0.155\\
$P_D$         &  0.082  &   0.107    &  0.077  &   0.041  &   0.012\\
$P_A$         &  0.012  &   0.033    &  0.004  &  $-$0.043  &  $-$0.088\\
$(1-P_B)$     &  $-$0.009 & $-$0.053 & $-$0.025 &  0.054 &   0.131 \\
\hline
\end{tabular}

\vspace{0.4 cm}
The first two lines of table 2, show the dependence on
the hadronic interaction lengths.
An increase in $\lambda_p$  decrease the   number of muons
because   with a larger proton interaction lengths
the shower develops deeper in the atmosphere  where the density is higher
and muons decay is more difficult.
When  $E_0/E_{\rm min}$ grows the effect of a changement of
$\lambda_p$   decreases in importance because a  growing fraction
the muons is produced
in  a cascade of type $p \to \pi \to \pi \to \mu$.

An increase of $\lambda_\pi$ increases the
number of  TeV muons  because if the interactions
length is longer the pions have  more time to
decay. The effect becomes smaller
with increasing $E_0/E_{\rm min}$, because   a larger $\lambda_\pi$
will also produce a deeper shower,
and the pions produced in a cascade of type  $p \to \pi \to \pi$
are created at lower altitude and  decay more rarely.
Note that the two  hadronic interaction lengths enter in the  expressions
for muon production only in the
combination: $\Lambda_\pi/\Lambda_p \simeq \lambda_\pi/\lambda_p$
and therefore $\xi(\lambda_\pi) \simeq - \xi(\lambda_p)$.

The other  lines  in table 2 describe the dependence on the differential
cross sections,  the same comments developed for the discussion of figures
2--5 apply,
one may notice the changements of sign of $\xi(K_p)$
(at very large $E_0/E_{\rm min}$) and of $\xi(P^*)$.
It is encouraging to see that  $|\xi_P|$  is always less than 1,
showing that the  sensitivity to  distortions of the spectral shape
is only moderate.
Note how   $G(x)$ is relatively insensitive  to the properties
of pion interactions.

\subsection {Fluctuations}

The  analytic method  that we have described  allows us to compute
the average  number of high energy muons in a shower,
but  does not take into account  fluctuations in the shower development.
In order to study  the importance of fluctuations and also to check the
analytic  calculation  we  have
prepared a  montecarlo implementation of
the family of interaction models discussed in section 3.
The straightforward montecarlo method is based of the shower code
developed in Bartol \cite{Wrotniak}  (see also \cite{Gaisser})
and is based on the following steps:
(i) a primary cosmic ray is  propagated in the atmosphere
until it interacts;
(ii) a set of secondary particles
(pions and nucleons) are produced at the interaction point
according to the  algorithms described  in section 3;
(iii)  each one  of the secondary particles is propagated
until it interacts or decay;
(iv) at each decay or interaction vertex  the incident particle
is  destroyed and a   set of new particles is produced
conserving energy and momentum;
(v) the procedure is  iterated until all
particles are below a preassigned minimum energy.
All produced muons are recorded.

In figure 6 we show the inclusive muon spectrum above 1 TeV
produced by a vertical primary proton of  energy 10, 10$^2$, $10^3$ and
10$^4$ TeV,  obtained with the montecarlo method and   with the  analytic
formula   using our `reference model'.
The results  obtained with the two methods are in excellent agreement
between each other.
The montecarlo method allows to study  not only $\langle N_\mu\rangle$
but also the probability distribution $P(N_\mu)$  of  having exactly
$N_\mu$ muons in a shower.
The distributions $P(N_\mu)$
calculated with the montecarlo method
for $E_0=10$ and $10^4$ TeV
($E_\mu^{\rm min} = 1$ TeV, $\theta = 0^\circ$ )
are shown in figure 7a and 7b.
In the same figures we  compare the montecarlo results
with a poissonian distribution of the same average
and a negative binomial distribution of same
average and dispersion. The  distribution $P(N_\mu)$ is  broader
than a poissonian, and the difference becomes more marked  with increasing
energy, the negative--binomial  being a good fit.
These results have been found previoulsy
by Forti and collaborators \cite {Forti}; we would  like to stress
that the non--poissonian fluctuations are not  connected to
violations  of Feynman or KNO scaling, and are  present also
in  exactly scaling models as those we  are discussing.

\section {Interaction Model and Multiplicity Distribution}
As an illustration of the importance of the interaction model for
the calculation of the multiplicity distribution of underground muons, we
have calculated the fluxes of TeV muons using a  `realistic' proton flux
of energy  spectrum: $\phi_0 (E) = 1.85~E^{-2.7}$,
steepening to $\propto E^{-3}$
for $E \ge 3 \times 10^{6}~GeV$
($E$ in GeV and $\phi$ in (cm$^2$ s sr)$^{-1}$) ),
and two different
models for the hadronci interactions.
In both models
$\lambda_p = 86$  and
$\lambda_\pi = 111.8$ g~cm$^{-2}$, and
the charged pion interactions:
are described by the algorithms originally proposed by Hillas
with parameters
$\{P_D,P_A, P_B \} = \{{1\over 2}, {1\over 2}, 0\}$.
The two models  differ in the treatment of
proton interactions.
The first model is nearly identical to the `reference model' with
parameters:
$\{K_p,P^*\} = \{ 0.475, 0\}$; the second model has  a larger
inelasticity but a softer spectrum:
$\{K_p,P^*\} = \{ {2 \over 3} , 1\}$.
The   $G$ functions obtained with the two models is
shown in figure 8, both models have $\epsilon_\mu(2.7) = 4.7 ~M_G(0.7) \simeq
8.16$ GeV, and  for energies $E \gg \varepsilon_\pi$
produce essentially identical `inclusive'
muon fluxes.
The larger multiplicity of  model--2 exactly compensates its softer spectrum.
We may however expect that using model--2
the probability of having several muons
in the same  shower is larger.

To investigate quantitatively this possibility
we have generated  approximately 3 million  vertical showers
with $E_0 \ge 1$ TeV for each of the two models. In order to  increase the
statistics of events with  high  muon multiplicity we have  sampled
the  energy of the showers from a  distribution
$\propto E_0^{-1.75}$, weighting each event
with $E^{1.75}/\phi_0(E)$.
In figure 9 we show the obtained inclusive muon fluxes that are
essentially identical in the two models, and
very well represented  by: $\phi_\mu (E) = \epsilon/E ~\phi_0(E)$
with $\epsilon = 8.16$ GeV.
The muon multiplicity   ($E_\mu \ge  1$ TeV)  for the two models
is shown in figure 10a, and the ratio  of the  fluxes obtained
the two models  for the same multiplicity is shown in  figure 10b.
The inclusive fluxes $\phi_1 + 2 \phi_2 + \ldots + n\phi_n + \ldots$
are equal to better that 1\%, but
the flux of single muons is
5.5\% smaller  using model-2,  the ratio
model--2/model--1 becomes 1.22 for double muons,
grows to 1.46 for triples, to 1.76 for quadruple muons, and then seems
to  remain approximately constant.

The two models considered are  `physically consistent',
both respect conservation laws (energy, momentum, baryon number),
and both produce the same inclusive muon flux,
however the  same experimental multiplicity distribution  of
underground muons if
interpreted with model--1 (model--2) would
result in  a  heavier (lighter) composition, because
the effects of the   smaller  (larger) frequency of
high multiplicity    events should be compensated  with a
different mass distribution of the primaries.

In this work we do not attempt a more realistic and complete discussion, that
will be presented  in a future paper.
We note that  the fairly  extreme distortions of the spectra
that we have  tried, produce   a difference of about a factor of 2
for  the frequency of events  with multiplicity $\aprge 10$.
The present range of  uncertainty in composition \cite{MACRO-composition}
can produce larger differences.

\section {Conclusions}
In this work we have  formally derived the
result that the average number of high energy   muons
($E_{\rm min} \aprge 1$ TeV) produced by a primary cosmic  ray proton
of energy $E_0$  has the
scaling form:
$\nmed =  G(E_{\rm min}/E_0)/E_{\rm min}$.
The function $G(x)$ is calculable analytically from  a knowledge of the
inclusive  single--particle  differential cross sections.
We have illustrated how the shape  and normalization
of $G(x)$ depends on the detailed form of these cross sections.

We do expect detectable   deviations from the  scaling behaviour.
A source of deviation is simply the fact that the
critical energy for kaon decay $\varepsilon_K$
is not small with respect to 1 TeV.
A second source  of deviation is due to the fact that
the hadronic interactions lengths are decreasing with
energy.
There is also the possibility of observable
violations of  Feynman scaling in the fragmentation region,
the measured growth of the  central plateau should have
visible effects for large $E_0/E_{\rm min}$.

We  have discussed a
possible generalization of the montecarlo algorithms
originally developed by Hillas \cite {Hillas},
to describe the properties of
of hadronic interactions.
These algorithms because of their remarkable simplicity and flexibility,
can be a useful too
to study in detail the effects of uncertainties in the
properties of hadronic interactions in the development of showers.
They  could be very useful  for the study of  the
highest energy cosmic rays ($E \sim 10^{20}$ eV).

Uncertainties in the modeling of hadronic interactions  are the dominant source
of systematic error for the  measurement  of   cosmic ray composition
from data on multiple muon events  in deep underground detectors.
The spectrum of the leading nucleon, and  of fast pions produced
in nucleon interactions are of  special importance.
The details of particle production in pion interactions are less important to
control.

\vspace {0.75 cm}

\noindent {\bf Acknowledgments}

Part of this work was developed during a visit to the
Bartol Research  Center whose hospitality is gratefully acknowledged.
Many  of the  ideas presented in this paper were formed
in conversations with T.K. Gaisser and T.Stanev.
I'm also grateful to Giuseppe Battistoni,
Sergio Petrera and Ornella Palamara for many
interesting discussions on the  problem of underground muons,
and to M.Hillas  for useful clarifications.
T.K Gaisser kindly read and commented  an early version of this paper.

\newpage
\noindent {\large \bf Figure Captions}

\vspace {0.3 cm}

\begin{itemize}

\item [] Fig. 1.
Curve $G(x)$ (equation (\ref{G-scal})) calculated  with the
Hillas model.
The curve is compared with the parametrization  of the average number of
muon at depth $h$ of Gaisser and Stanev \cite{NIM85} and Forti {\it et al.}
\cite {Forti}
assuming the  approximate  relation $E_{\min} = 0.53 ~(e^{0.4 h}-1)$
($E$ in TeV, $h$ in km.w.e.).

\item [] Fig. 2a. Plot  of the inclusive ($p \to \pi^\pm$)  spectrum
calculated with  $P^* = 0$ and  with inelasticity
$K_p = {1 \over 2}$,  ${1 \over 3}$,  ${2 \over 3}$.

\item [] Fig. 2b. Curve $G(x)$ calculated   with
the `reference model'
[$\{K_p,P^*;~ P_D, P_A, P_B\}
= \{ {1\over 2}, 0, {1\over 2}, {1\over 2}, 0\}$,
$\lambda_p = 86$, $\lambda_\pi = 111.8$ g cm$^{-2}$]
 and  with   modified inelasticity
$K_p = {1 \over 3}$,  ${2 \over 3}$.

\item [] Fig. 3a. Plot  of the inclusive ($p \to \pi^\pm$)  spectrum
calculated with
$P^* = 0$ and  $P^* = 1$ ($K_p = {1 \over 2}$).
{}.

\item [] Fig. 3b. Curve $G(x)$ calculated   with the
`reference model' ($P^* = 1$)  and with the modification
$P^*= 1$.

\item [] Fig. 4a. Plot  of the inclusive ($\pi^\pm \to \pi^\pm$)  spectrum
calculated with  the   model
$\{P_D,P_A, P_B \} = \{P_D, {1\over 2}, 0 \} $ and
$P_D = 0$, ${1\over 2}$, 1.

\item [] Fig. 4b. Curve $G(x)$ calculated   with the `reference model'
($P_D = {1\over 2}$) and  with $P_D = 0$, 1.

\item [] Fig. 5a. Plot  of the inclusive ($\pi^\pm \to \pi^\pm$)  spectrum
calculated with  the   model
$\{P_D,P_A, P_B\} = \{ {1\over 2},{1 \over 2}, P_B \}$ and
$P_B = 0$, 1.

\item [] Fig. 5b. Curve $G(x)$ calculated   with the `reference model'
($P_B =0$) and with $P_B = 1$.

\item [] Fig. 6.
Plot of $f(E_\mu; E_0)$  the inclusive differential  muon spectrum
produced by a vertical primary proton  for $E_0 = 10$, 100, $10^3$ and
10$^4$ TeV. The interaction model used is the `reference model'.
The curves are analytic calculations, the histograms the results
of  Montecarlo runs.

\item [] Fig. 7a.
Probability distribution  $P(N_\mu; ~E_0, E_{\rm min})$
that a primary proton of energy $E_0$ produces
$N_\mu$ muons above  threshold energy $E_{\rm min}$.
The points are the results of Montecarlo  calculation
with $\theta = 0^\circ$, $E_{\rm min} = 1$ TeV, $E_0 = 10^2$ TeV.
The  dashed curve is  a poissonian distribution with the same
average, the solid line is a
negative--binomial  distribution of same average  and dispersion,
as the montecarlo result.

\item [] Fig. 7b.
As in figure 7a,  $E_0 = 10^4$ TeV.

\item [] Fig. 8.
Function G(x) for two models
that yield the same inclusive muon distribution.
The proton interactions are described by
$\{K_p,P^*\} = \{ 0.475, 0\}$  in model--1 and
$\{ {2 \over 3} , 1\}$ in model--2.
The other parameters are chosen as in the `reference model'.

\item [] Fig. 9.
Inclusive muon flux calculated with analytic and montecarlo methods
assuming the   primary proton flux
$\phi_0 (E) = 1.85~E^{-2.7}$
 steepening to $\propto E^{-3}$
for $E \ge 3 \times 10^{6}~GeV$.
(see text).

\item [] Fig. 10a.
Muon multiplicity distribution  $\phi_\mu(N_\mu)$
($E_{\rm min} = 1$ TeV and $\theta = 0^\circ$ )
calculated with a montecarlo method
assuming a primary cosmic ray flux of protons
with  energy  spectrum: $\phi_0 (E) = 1.85~E^{-2.7}$
 steepening to $\propto E^{-3}$
for $E \ge 3 \times 10^{6}~GeV$.
The two set of points refer to two different proton
interaction models :
$\{K_p,P^*\} = \{ 0.475, 0\}$
and $\{K_p,P^*\} = \{ {2 \over 3} , 1\}$ that result in the same inclusive
muon distribution.

\item [] Fig. 10b.
Ratio of the fluxes $\phi_\mu(N_\mu)$ of figure 9a, calculated with two
different models for  proton iteractions.

\end{itemize}

\end{document}